\documentclass[usenatbib,useAMS,preprint,psfig2]{mn2e}
\usepackage{times}
\usepackage{amsmath}
\usepackage{amssymb,amsfonts}
\usepackage{graphicx}
\usepackage{epstopdf}
\usepackage{verbatim}

\def\gapprox{\lower.4ex\hbox{$\;\buildrel >\over{\scriptstyle\sim}\;$}}
\def\lapprox{\lower.4ex\hbox{$\;\buildrel <\over{\scriptstyle\sim}\;$}} \def\be{\begin{equation}}
\def\be{\begin{equation}}
\def\ee{\end{equation}}
\def\bea{\begin{eqnarray}}
\def\eea{\end{eqnarray}}
\catcode`\@=11
\font\tenmib=cmmib10
\font\tensyb=cmbsy10
\font\tenbi=cmmib10 
\newfam\bifam  
\textfont\bifam=\tenbi  

\def\unboldmath{\everymath{}\everydisplay{}
          \textfont\@ne\teni 
          \textfont\tw@\tensy
          }
\def\boldmath{$\!\!$\relax\everymath{\mit}\everydisplay{\mit}
        \textfont\@ne\tenmib
        \textfont\tw@\tensyb 
        \relax}%
\catcode`\@=12
  
\begin{document}
\title[Interstellar Holography]{Interstellar Holography}
\author[Walker, Koopmans, Stinebring \&\ van~Straten]{M.A.~Walker$^{1,2,3,4}$, L.V.E.~Koopmans$^3$, D.R.~Stinebring$^{5,6}$, W.~van~Straten$^{4,7,8}$\\
1. Manly Astrophysics Workshop Pty Ltd, Unit 3, 22 Cliff Street, Manly, NSW 2095, Australia\\
2. School of Physics, University of Sydney, NSW 2006, Australia\\
3. Kapteyn Astronomical Institute, University of Groningen, P.O. Box 800, 9700 AV Groningen,
The Netherlands\\
4. Netherlands Foundation for Research in Astronomy, P.O. Box 2, 7990 AA Dwingeloo, The Netherlands\\
5. Oberlin College, Department of Physics and Astronomy, Oberlin, OH 44074, U.S.A.\\
6. Leiden University Observatory, Leiden, The Netherlands\\
7. Center for Gravitational Wave Astronomy, University of Texas, Brownsville, TX 78520, U.S.A.\\
8. Swinburne University of Technology, Centre for Astrophysics and Supercomputing, Hawthorn, VIC 3122, Australia}

\date{\today}

\maketitle

\begin{abstract}
The dynamic spectrum of a radio pulsar is an in-line digital hologram of the ionised interstellar medium. It has previously been demonstrated that such holograms permit image reconstruction, in the sense that one can determine an approximation to the complex electric field values as a function of Doppler-shift and delay, but to date the quality of the reconstructions has been poor. Here we report a substantial improvement in the method which we have achieved by simultaneous optimisation of the thousands of coefficients that describe the electric field.  For our test spectrum of PSR B0834+06 we find that the model provides an accurate representation of the data over the full 63~dB dynamic range of the observations: residual differences between model and data are noise-like. The advent of interstellar holography enables detailed quantitative investigation of the interstellar radio-wave propagation paths for a given pulsar at each epoch of observation; we illustrate this using our test data which show the scattering material to be structured and highly anisotropic. The temporal response of the medium exhibits a scattering tail out to beyond $100\,{\rm\mu s}$ and a pulse arrival time measurement at this frequency and this epoch of observation would be affected by a mean delay of $15\,{\rm\mu s}$ due to multipath propagation in the interstellar medium.
\end{abstract}

\begin{keywords}
techniques: interferometric ---  pulsars: general --- pulsars: individual: B0834+06 --- scattering --- ISM: structure --- turbulence
\end{keywords}

\section{Introduction}\label{sec:intro}

With modern instrumentation pulsar dynamic spectra can be recorded at high spectral and temporal resolution yielding a dataset with a large information content from just one observation. For example: if we observe a pulsar for one hour, sampling the spectrum with 1~kHz channels every 10 seconds, over a total bandwidth of 100~MHz, then we have $\sim4\times10^7$ independent flux measurements. And if the pulsar is bright and the telescope is large then each of these measurements can have signal-to-noise ratio of order unity implying a total information content of potentially $\sim$40~Mbit. This information relates primarily to multipath scattering of the radio-waves by the ionised Interstellar Medium (ISM), as it is this scattering which gives rise to the observed interference fringes.  It may be that these paths are determined by random irregularities -- e.g.  caused by turbulence -- in the ISM. In that case any given dynamic spectrum contains information about those random irregularities, and there is no strong motivation to study the individual spectra in detail as they reflect particular realisations of a stochastic process.   But some pulsar dynamic spectra exhibit a high level of order in their fringe patterns --- see Rickett (1991) for an overview. This fact has been emphasised by consideration of the two-dimensional power spectra of the dynamic spectra, wherein power is often seen to be concentrated in parabolic arcs and inverted arclets (Stinebring et al 2001). There is a consensus that this phenomenon arises directly from the geometry of the scattering process (Stinebring et al 2001; Cordes et al 2006;  Walker et al 2004), with waves scattered through an angle $\vec{\theta}$ experiencing a Doppler-shift proportional to one component of $\vec{\theta}$ and a delay proportional to $\vec{\theta}\cdot\vec{\theta}$. In cases where the parabolae are very sharp it has been argued that the scattering is highly anisotropic (Walker et al 2004; Cordes et al 2006). Sharply defined arcs/arclets also require that the scattering arises in a region of small extent along the line-of-sight (Stinebring et al 2001), so it is not distributed turbulence but discrete, localised  structures which are responsible for these features. However, the physical nature of the scattering media remains obscure. Consequently there is now some incentive to explore the information content in individual dynamic spectra, in order to build a detailed picture of the scattering structures. Further motivation for investigating  individual dynamic spectra is provided by studies of the pulsars themselves: if the properties of the scattering screen are accurately known it is possible to use the screen for very high resolution imaging of the pulsar magnetosphere (Wolszczan and Cordes 1987; Gwinn et al 1997; Walker and Stinebring 2005, WS05 henceforth). Precision pulsar timing programs (e.g. Manchester 2007) also provide an incentive for understanding the particular interstellar propagation paths which contribute to individual observations --- if the propagation delays remain uncorrected in the data they constitute a potentially large systematic error in pulse arrival time measurements. 

It has previously been demonstrated that  one can iteratively construct a model of the electric field as a function of radio-frequency and time, starting from the observed dynamic spectrum (WS05). The procedure for achieving this is largely equivalent to determining a phase for the electric field in each pixel of the dynamic spectrum, because a noisy estimate of the field amplitude is given directly by forming the square root of the observed intensity. This situation is common to the broad class of problems known as ``phase-retrieval problems'', which are well known in the optics literature (e.g. Fienup 1982). However, the method of solution demonstrated for pulsar dynamic spectra appears to be new; it exploits sparseness of the power distribution in the Fourier domain, and a solution is obtained iteratively by adding discrete new field components in such a way as to minimise the differences between the model dynamic spectrum and the data. Conceptually the process has a strong similarity to the CLEAN algorithm (H\"ogbom 1974) which is commonly used in  radio astronomical imaging for deconvolving the synthesised telescope beam from a ``dirty'' image of the sky; CLEAN usually works well if the image is only sparsely populated with emission. The connection between the two algorithms is reinforced when we recognise that determining the electric field from the dynamic spectrum is also equivalent to a deconvolution. The dynamic spectrum as a function of radio-frequency and time is simply $I(\nu, t) = U^*(\nu, t)\,U(\nu, t)$, where $U$ is the electric field; in the Fourier domain this relationship becomes a convolution $\widetilde{I} = \widetilde{U}^*\otimes\widetilde{U}$ and so we are deconvolving the Fourier transform of the electric field from its complex conjugate.

Some comments about nomenclature are appropriate at this point. Any process which allows us to reconstruct the electric field which gave rise to a recorded fringe pattern is sensibly termed ``holography'', and the recorded fringe pattern (i.e. the dynamic spectrum in our case) is the hologram. In the case considered here it is ``digital holography'' because the reconstruction is done in software, and ``in-line'' because the object (i.e. the scattering screen in our case) is transparent and sits in the beam which forms the reference wave. In-line holography is sometimes called ``Gabor holography'' because it is the arrangement which was originally conceived by the inventor of holography, Dennis Gabor. 

To date the fidelity of electric field reconstructions from pulsar dynamic spectra has been poor. Although the model of WS05 successfully reproduces the general appearance of their test dynamic spectrum there are substantial quantitative differences between the model and the data. These errors are seen when the model ``secondary spectrum'' (i.e the power spectrum of the dynamic spectrum) is compared to the data: the data exhibit a dynamic range of 63~dB, but the {\it difference\/} between the WS05 model secondary spectrum and the secondary spectrum of the data has a dynamic range of 47~dB, implying that the reconstruction has correctly captured only the top 16~dB of the data. Here we report modifications to the reconstruction process which have yielded dramatic improvements to the accuracy of the electric field model; on the same test dataset as used by WS05 we find that our improved model captures the full dynamic range of the data. These gains  were achieved by simultaneous optimisation of the thousands of parameters describing the wave interference phenomenon, and by simultaneous optimisation of the hundreds of parameters which describe the intrinsic temporal flux variations of the pulsar.

This paper is organised as follows: in the next section we detail the improvements we have made to the reconstruction algorithm described by WS05; in \S3 we present the results we have obtained, using the same test data employed by WS05; and in \S4 we consider what our test data tell us about the interstellar medium, with emphasis on the temporal response of the scattering medium.

\section{Optimisation of the E-field model}
In WS05 we described an algorithm for modelling the electric field structure in the Fourier domain conjugate to the dynamic spectrum. The latter is recorded as a function of radio-frequency, $\nu$, and time, $t$, and the corresponding conjugate variables are delay, $\tau$, and Doppler-shift, $\omega$.  The WS05 algorithm proceeds from a grid of noisy measurements of the electric field envelope, $|U(\nu,t)|^2$, to a model of $\widetilde{U}(\tau, \omega)$ in terms of discrete wave components, $j$:
\be
\widetilde{U}(\tau, \omega) = \sum_j \tilde{u}_j\,\delta(\tau-\tau_j)\,\delta(\omega-\omega_j),
\ee
and the components (characterised by $\tau_j$, $\omega_j$ and the complex number $\tilde{u}_j$) are chosen one-by-one so as to minimise the differences between model and data.  With $\sim9000$ components the model reported by WS05 gives a good visual match to the data, but the residuals are large in comparison with the noise level, implying large systematic errors.  An inspection of the differences between the observed secondary spectrum (i.e. the power spectrum of the dynamic spectrum) and the model secondary spectrum -- the two quantities shown in  the lower panel of figure 1 in WS05 --  reveals that the discrepancies occur predominantly at the same locations in delay-Doppler space where there is already power present in both model and data. This suggests that the systematic errors introduced by the WS05 algorithm are not due to incorrectly placed components (i.e. wrong $\tau_j, \omega_j$), or an insufficient number of components in the model, but rather due to errors in determining the various $\tilde{u}_j$.

It was noted in WS05 that global optimisation of the $\{\tilde{u}_j\}$ is desirable, in order to reduce systematic errors in the model, but difficult to achieve because of the large number of free parameters which would be involved in such an optimisation.  In particular inversion of a $10^4\times10^4$ non-sparse matrix  -- which is perhaps the most obvious approach to solution of the linearised least-squares optimisation -- is computationally challenging. Furthermore any approach based on solution of the linearised problem would require iteration in order to solve the full non-linear optimisation problem.  Fortunately there are easier methods -- see, for example, Nocedal and Wright (1999) -- and we have used one of these to optimise the WS05 electric field model as we now describe.

The method which we employed is a quasi-Newton method, in which a demerit function $S$ is  minimised by seeking successively closer approximations to the solution of $\partial S/\partial x_i=0$ for all parameters $x_i$ over which we wish to optimise. Newton's method requires knowledge of the Hessian (i.e. all the second derivatives $\partial^2 S/\partial x_i\partial x_j$), which is computationally expensive when a large number, $N$, of parameters is involved. By contrast, quasi-Newton methods proceed by approximating the Hessian; information from current and previous iterations is used to update the approximate Hessian for subsequent iterations yielding, in effect, a finite-difference representation of the local curvature of  $S$. The update scheme which we used is the BFGS (Broyden-Fletcher-Goldfarb-Shanno) update. Specifically: we used the L-BFGS-B algorithm (Byrd et al 1995), which is a ``Limited memory'' implementation of BFGS, using line-search minimisation, in which Bounds are permitted on the parameters $x_i$. The term ``Limited memory'' here refers to the fact that the $N\times N$ Hessian is replaced by the outer product of two $m\times N$ matrices, with $m$ being the number of prior iterations which are employed in constructing the update. Because $m\sim\;$a~few, and $m$ does not grow with $N$, the storage requirements of the algorithm are only modest and grow linearly with $N$. The L-BFGS-B code was written by specialists in the field of numerical optimisation; it is freely available as a set of FORTRAN subroutines\footnote{{\tt http://www.ece.northwestern.edu/$\sim$nocedal}}.  In order to make use of this code it is necessary for the user to supply routines which evaluate the demerit function, $S$, and its  partial derivatives, $\partial S/\partial x_i$, with respect to all the parameters, $x_i$, over which we wish to  optimise. Given these inputs the L-BFGS-B code will search for a minimum in $S$. If a minimum is found by L-BFGS-B it is not guaranteed to be a global minimum.  The L-BFGS-B package is well documented; there are clear instructions on how to use the code and simple drivers are included in the package so that the user can verify correct performance on their own computer. For our application we used a model with two sets of parameters: the various $\tilde{u}_j$, each of which is described by two unbounded real variables, representing the real and imaginary parts, and a set of positive-definite real numbers describing the intrinsic pulsar flux as a function of time, $\{f_k\}$. The inclusion of the various parameters $f_k:\;f_k\ge0\;\;\forall\;\; k$ demands that the optimisation software be able to handle variables which are bounded, as is the case for the L-BFGS-B package. The WS05 algorithm does not attempt to solve for the $f_k$ explicitly but simply applies a Fourier-domain filter to the data in an attempt to remove the intrinsic pulsar flux variations. Because there is no clear-cut distinction between intrinsic flux variations and those due to wave interference, the procedure used by WS05 is quite crude and in the present work we use it only as a starting point (as per item (i) in \S3.1 of WS05).

Our first attempt at improving on the WS05 approach was to take the output of the WS05 algorithm and use it as the starting point of an optimisation with L-BFGS-B. The results were good, yielding a model which captured much more of the dynamic range in the data, and had a lower value of the reduced $\chi^2$ statistic. This result was encouraging, but it was clear that we could do better: the components which are identified at each iteration in the WS05 algorithm depend on the electric field model at that point, so the systematic errors in the WS05 model are not completely eliminated by {\it post facto\/} optimisation --- spurious components remain in the optimised model, albeit at a low level.  If, on the other hand, the electric field model at each iteration of the WS05 algorithm is optimised (using L-BFGS-B) then these spurious features can be minimised. This approach has the additional benefit of simplicity in the processing of the data, requiring only one pass through the data and one set of code. We therefore implemented a new iterative decomposition algorithm, in FORTRAN, which employs the L-BFGS-B package to optimise the model $\tilde{u}_j$ and $f_k$. The new algorithm differs from WS05 in the following ways:

\begin{enumerate}

\item The parameters $f_k$ (describing the intrinsic pulsar fluxes) are optimised once for every 100 new field components which are picked. The optimisation over $\{f_k\}$ is done separately from that over $\{\tilde{u}_j\}$.

\item The number of new electric field components picked at each iteration is given by the integer part of $1+N_c/100$, where $N_c$ is the current number of electric field components. So initially only one new component is picked at each iteration, and when the model contains a large number of components the fractional increase per iteration is 1\%. 

\item When the number of electric field components exceeds 100, the algorithm is free to pick components which have $\tau<0$. 

\item The algorithm terminates when the reduced $\chi^2$ (i.e. $\chi^2$ per degree of freedom) reaches unity, or when the reduced $\chi^2$ reaches a minimum --- whichever occurs first.

 \end{enumerate}
Because of the high dynamic range of the data it is important to maintain high precision in the optimisation, so we set the L-BFGS-B parameter named ``factr'' (which measures the precision in units of machine precision) to the value 10. The number of previous steps used in forming each BFGS update is $m=10$. 

\begin{figure*}
\includegraphics[width=15cm]{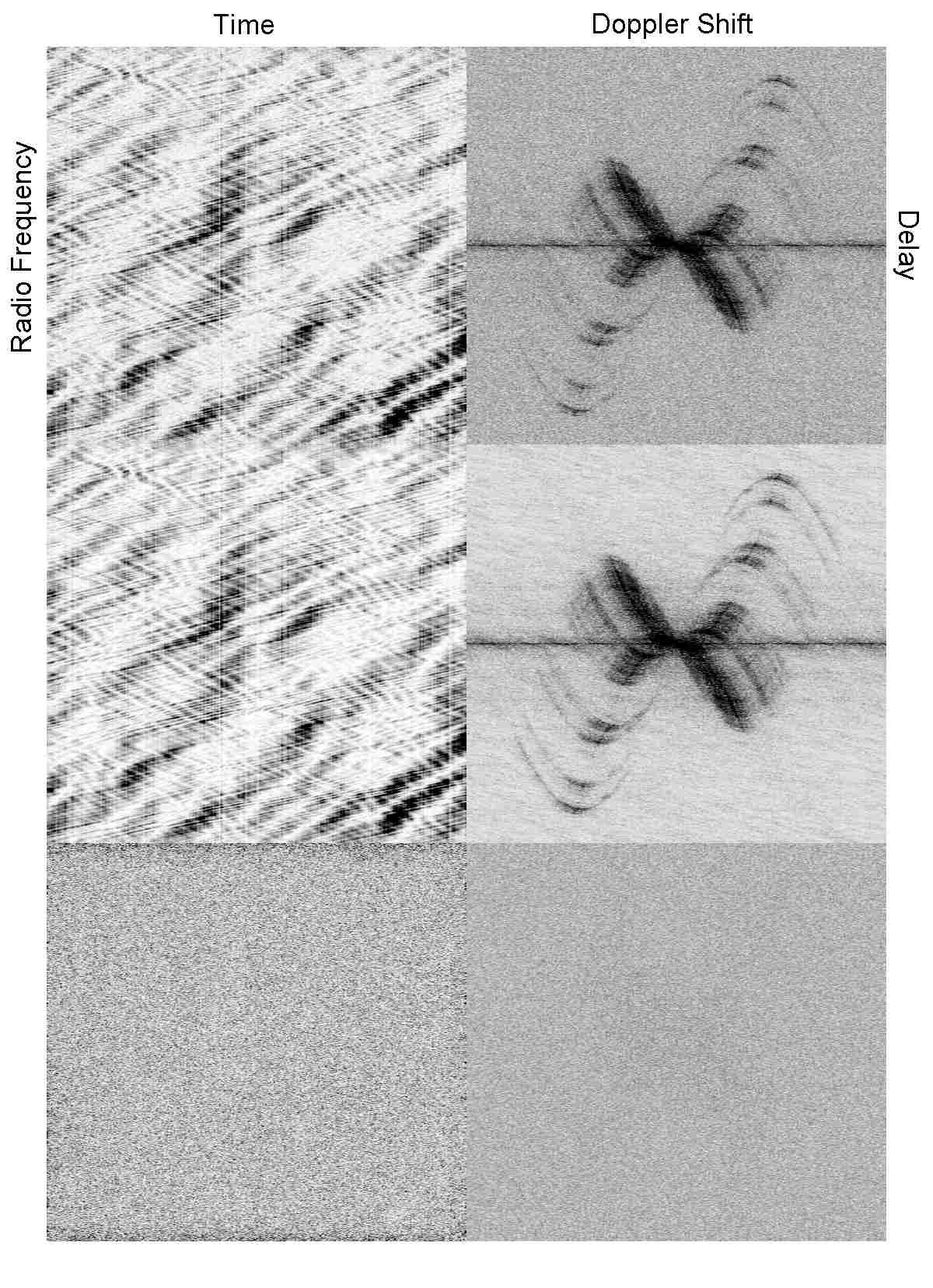}
\caption{Data (top), model (middle), and residuals between data and model (bottom), for an observation of PSR~B0834+06 in a 1.56~MHz band centred on 321.00~MHz. The data were taken with the Arecibo radio telescope in conjunction with the WAPP backend signal processing units on MJD~53009; there are 1024 spectral channels, and 270 time samples each of 10 seconds duration. The left-hand panels show dynamic spectra, while the right-hand panels show the corresponding secondary spectra (power spectra of the dynamic spectra); the range of the secondary spectra is $\pm50$~mHz in Doppler-shift and $\pm327.6\;{\rm \mu s}$ in delay. Inverse grey-scale (black is peak intensity) is used in all cases; the transfer function is linear for the dynamic spectra, and logarithmic for the secondary spectra. All secondary spectra shown here have the same transfer function; the transfer functions for dynamic spectra are identical in the case of data and model, whereas the output range is stretched by a factor of five for the dynamic spectrum residuals in order to reveal their structure. In comparison with figure 1 of WS05 note that the present figure includes the modulating effects of the intrinsic pulsar flux variations. We have chosen to display the full secondary spectra, including negative delays, even though these spectra are symmetric through the origin, so that the structure near zero delay can be better seen.}
\end{figure*}

\section{Results}
A direct comparison of the results of WS05 with the new algorithm is possible by using the same test data as WS05 employed. Those data are shown in figure 1, along with the model generated by the new algorithm; both model and data are shown in the form of a dynamic spectrum and its power spectrum (the ``secondary spectrum''). The model was generated using the algorithm described in \S2. Figure 1 also shows the residuals between data and model dynamic spectra, and the power spectrum of those residuals; the residuals appear noise-like in both panels. Some quantitative measures of the success of the new algorithm are appropriate. The new algorithm achieves a reduced chi-squared value of $\chi^2_r= 1.00$ (this was the stopping criterion which was reached first), versus $\chi^2_r=1.19$ achieved by WS05, and it does so with only 8,000 electric field components versus 8,720 in WS05. Note, however, that the new algorithm does employ an additional 270 real numbers -- one for each time sample -- to describe the intrinsic pulsar flux variation with time; these numbers were in effect fixed in the WS05 algorithm by a Fourier-plane filter acting on the input data. Subtracting the model dynamic spectrum from the data, and then forming the power spectrum of the residuals gives a sensitive test of the fidelity of the model because it picks out correlated errors in the dynamic spectrum model. For the new algorithm the largest value of the residual power is only 11~dB above the mean noise power in the data, whereas the peak power in both data and model is 63~dB above the mean noise power in the data. Thus the algorithm is certainly free of systematic errors over a range of  52~dB.

In fact the new algorithm has achieved noise-limited performance and is capturing the full dynamic range of the data; to see this we need only examine the statistics of the noise power, shown as a histogram in figure 2. If the residuals were purely noise-like then the expected probability distribution function would be an exponential, because the residual power is the sum of the squares of two independent variables each of which has the same Gaussian distribution. We can see from figure 2 that the residuals conform closely to this expectation; we can also see that the peak residual is not introduced by any systematic error in the modelling but rather it is simply the tail of the noise distribution. The residuals actually exhibit a slightly lower noise level than the data (dashed line); this can be understood by reference to figure 1. The noise in the data is estimated from the floor power level in the secondary spectrum, specifically an area in the corner of the secondary spectrum (away from any obvious signal power) is chosen and the mean power over this area is computed. The model secondary spectrum also exhibits a floor power level, even though the model is intended to represent ``signal'' rather than ``noise''; this power is not present in the residual dynamic spectrum so the mean power level in the residuals is slightly lower than in the data. 

The model electric field strength determined by our new algorithm is shown in figure 3. Visually this result is similar to that obtained by WS05 (their figure 2), with the notable exception that in WS05 all field amplitudes were fixed at  zero in the region $\tau<0$ (so that region is not displayed in their figure 2). On physical grounds this area is expected to be free of astronomical signals. However, in-line holography generally involves a certain amount of confusion between the image and its conjugate, because the observable quantity is usually the intensity $U^*U$ which is invariant under the replacement $U\rightarrow U^*$. In the present context the conjugate image appears at negative delays, because forming the complex conjugate of $\exp[ 2\pi i (\nu\tau_j + \omega_jt]$ is equivalent to making the replacements $\tau_j\rightarrow-\tau_j$,  $\omega_j\rightarrow-\omega_j$. In our figure 3 it can be seen that there is little  trace of an inverted parabola in the lower half of the figure -- only weak components can be seen in the region $\tau<0$ -- indicating minor confusion between the image and its conjugate (see also \S4.1). This clean separation is another indication of the low level of systematic errors inherent in the new algorithm. In the same vein we note that there is little power evident in figure 3 near the $\tau=0$ locus (a horizontal line tangent to the apex of the parabola), indicating that the intrinsic pulsar flux variations, described by the various $f_k$, have been accurately modelled.

Although the origin of coordinates (i.e. $\tau=0, \omega=0$) in figure 3 is in principle unknown (WS05), in the case of this particular dataset it appears sensible to assign the origin to the centre of the image because this is where the largest amplitude field component is located and the great majority of the intensity in figure 3 lies above this point --- consistent with the brightest image component being an unscattered wave.

\begin{figure}
\vskip -1.5cm\hskip -1.5cm\includegraphics[width=11cm]{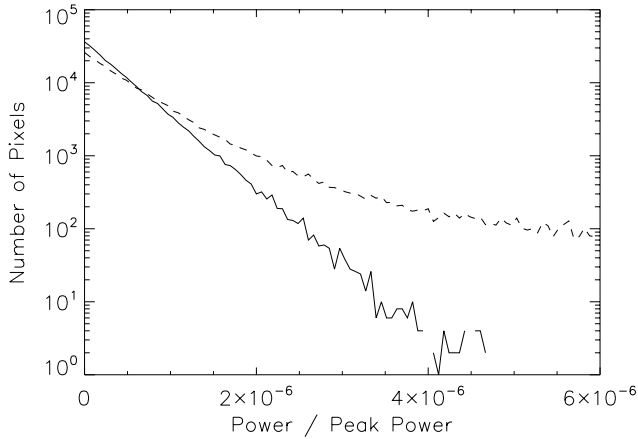}
\vskip -6.5cm
\caption{Histogram of the number of pixels per bin (of width $6\times10^{-8}$) as a function of power, for the two-dimensional power spectrum of the residual between the model dynamic spectrum and the data (solid line). Also shown is the corresponding histogram for the data (dashed line). The residual power distribution corresponds very closely to the expected exponential noise distribution. The mean power in the residuals is slightly less than that calculated from the data because the latter estimate includes a small contribution from signals which are incorporated into the model.}
\end{figure}

\begin{figure}
\includegraphics[width=8.5cm]{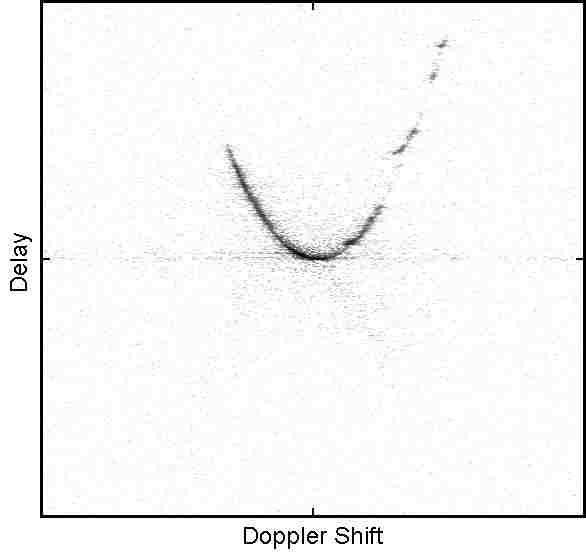}
\caption{Model electric field amplitudes corresponding to the model spectra shown in figure 1. Here the amplitudes are shown in grey-scale, as a function of Doppler-shift and delay, with a logarithmic transfer function; peak intensity is black. The Doppler-shift  ranges over a total of 100~mHz (270 pixels) and the delay  ranges over a total of $655\;{\rm \mu s}$ (1024 pixels). Absolute values of delay and Doppler-shift are in principle unknown but it appears sensible to assign the origin of coordinates to the centre of the image (which is the peak field amplitude) in this particular case. It is evident from this figure that the image has been well separated from its complex conjugate, which would appear as an inverted parabola. Also notable is the low level of power around zero delay,  which indicates that the intrinsic pulsar flux modulations have been accurately determined by the algorithm.}
\end{figure}

Having now reached the point where we can form models which are a good match to the data it is appropriate to draw some inferences about the properties of the scattering medium which gives rise to the data shown in figure 1. Excepting the temporal analysis in \S4.1 our discussion is only  qualitative; that is because we have constructed our holographic image in delay-Doppler space (i.e. the Fourier space conjugate to the frequency-time space in which the data are recorded), whereas progress in understanding the scattering medium relies on a knowledge of the image in two-dimensional spatial (angular) coordinates and there is no simple way of proceeding from the former to the latter. 

\section{Properties of the scattering medium}
An important qualitative aspect of figure 3 is that power in the model electric field is tightly concentrated around a parabolic locus, with delay proportional to the square of the Doppler-shift; this much was already evident in WS05. This confirms that the underlying scattering is highly anisotropic, with a scattered image which is very much longer than it is wide --- a conclusion which has previously been arrived at from consideration of the properties of observed secondary spectra for this and other pulsars (Walker et al 2004; Cordes et al 2006; Trang and Rickett 2007). 

We can also see from figure 3 that some parts of the parabola show high intensity levels while others are almost completely devoid of signal, and in some places the high-low transitions are fairly abrupt. Abrupt changes are suggestive of well-defined boundaries to the scattering regions. Four intensity concentrations are seen at large delay and positive Doppler-shift; these correspond to the features named ``A,B,C,D'' by Hill et al (2005) in a secondary spectrum analysis of data spanning more than three weeks. These features are presumably due to localised plasma concentrations; it is not yet clear whether these concentrations are diffracting or refracting the radio-waves into the telescope. Feature ``A'' lies significantly above the locus of the parabola; this extra delay could be the wave-speed (dispersive) delay of a high column-density structure --- an interpretation which can in future be tested by comparing data at two different frequencies obtained at the same epoch.

\begin{figure}
\vskip -1.5cm\hskip -1.5cm\includegraphics[width=11cm]{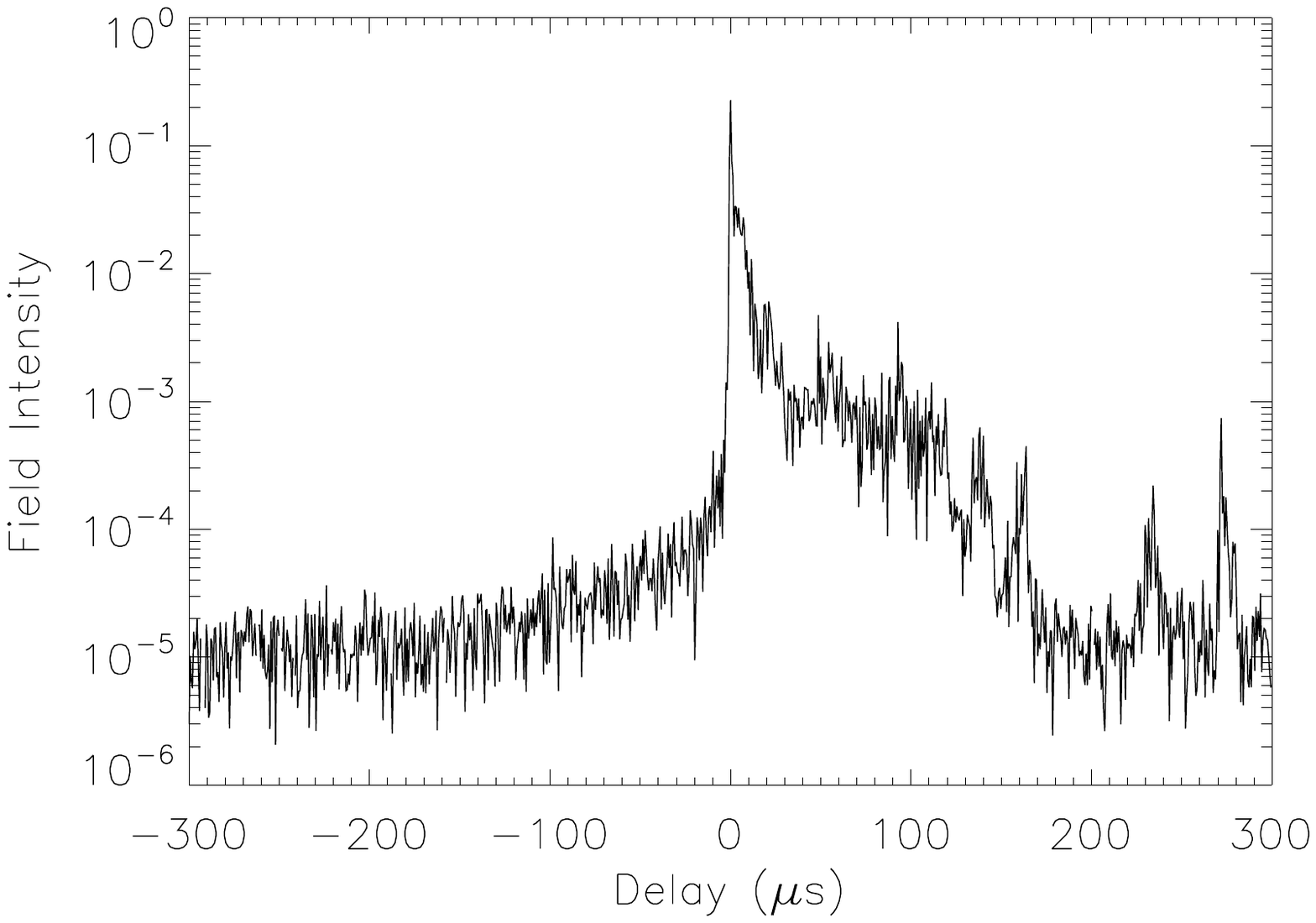}\vskip -6.5cm
\vskip -1.5cm\hskip -1.5cm\includegraphics[width=11cm]{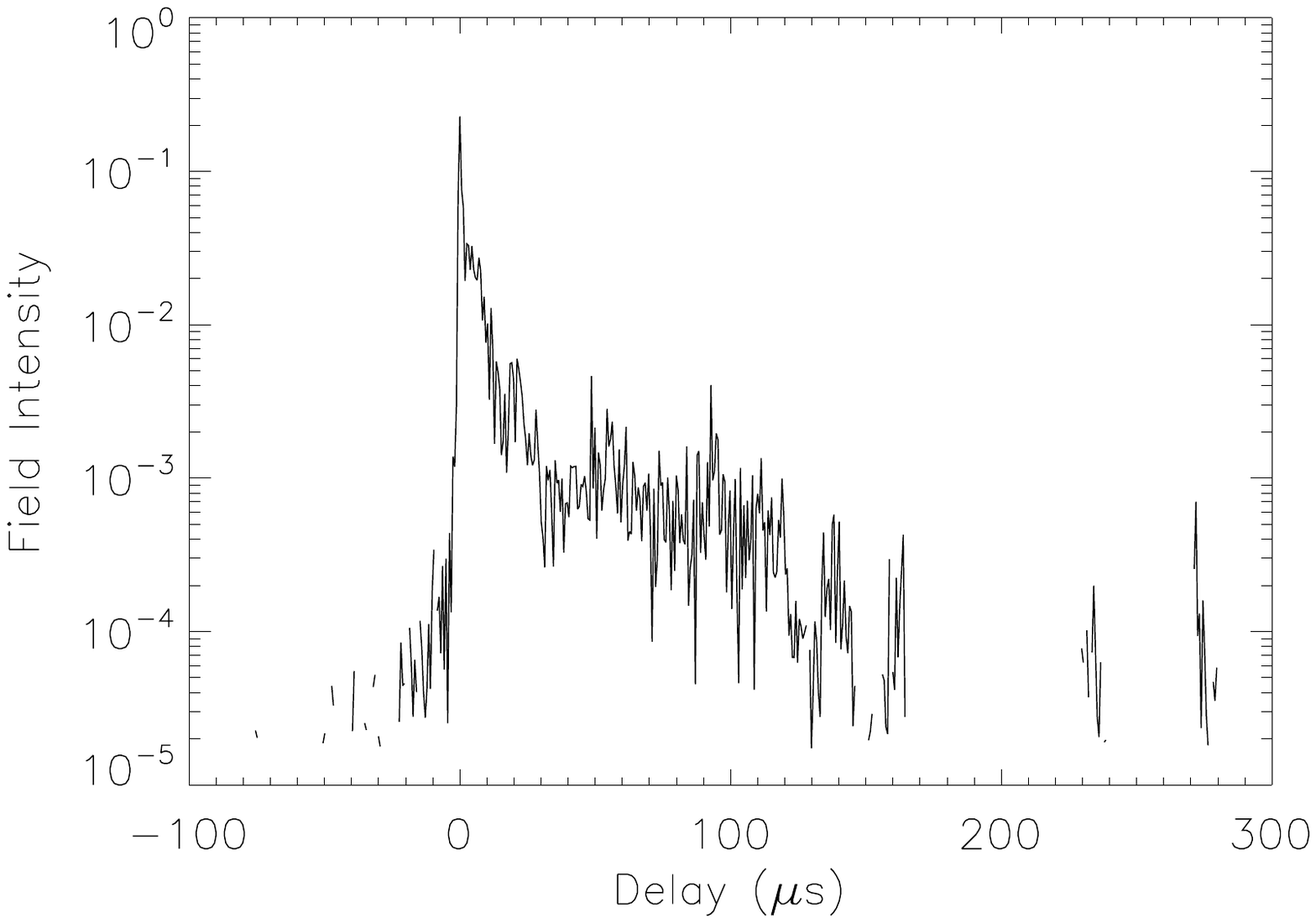}\vskip -6.5cm
\caption{The mean intensity impulse response of the scattering medium: top panel, as determined from the holographic image shown in figure 3; bottom panel, as determined from the holographic image shown in figure 3 with low amplitude coefficients ($|\tilde{u}_k|<0.004$) set to zero.}
\end{figure}

\subsection{Temporal response of the medium}
Inferring the spatial structure of the scattering medium from our image of  $\widetilde{U}$ is not a simple exercise and it is beyond the scope of this paper to attempt an in-depth analysis. On the other hand the holographic image shown in figure 3 is well suited to determining the temporal response of the medium which is introduced by multi-path propagation. The electric field amplitude as a function of delay, $\tau$, and observing time, $t$, can be determined by forming the inverse Fourier transform of equation 1 with respect to  $\omega$:
\be
{\cal U}(\tau, t) = \sum_j \widetilde{u}_j\,\delta(\tau-\tau_j)\exp\left[ 2\pi i \omega_jt \right]
\ee
(cf. equations 1 and 2 of WS05). The corresponding field intensity is
\be
{\cal I}(\tau, t) = {\cal U}^* {\cal U} = {\cal A}(\tau)\;+ \;{\cal B}(\tau, t),
\ee
where
\be
{\cal B}(\tau, t):=\sum_{j\ne k}\widetilde{u}_j^* \widetilde{u}_k\,\delta(\tau-\tau_j)\,\delta(\tau-\tau_k)\exp\left[ 2\pi i (\omega_k-\omega_j)t \right],
\ee
describes the beating between waves of differing Doppler-shifts (but the same delay) and
\begin{eqnarray}
{\cal A}(\tau) := \sum_j\widetilde{u}_j^* \widetilde{u}_j\,\delta(\tau-\tau_j)\cr
=  \int\!\!{\rm d}\omega\;\left|\widetilde{U}(\tau, \omega)\right|^2,\;\,
\end{eqnarray}
is the mean intensity impulse response function of the scattering medium.  The function ${\cal A}(\tau)$ convolved with the intrinsic pulse profile yields (up to a normalisation factor) the average pulse profile for this observation. For our image, $\widetilde{U}$, the mean intensity impulse response, ${\cal A}$,  is shown in the top panel of figure 4. For those delays where $\widetilde{U}$ includes one wave component which has a much larger amplitude than the other components at that delay the beat terms will all be relatively small and $|{\cal B}|$ will be small compared with ${\cal A}$. In general, however,  ${\cal B}$ is not negligible; we defer consideration of ${\cal B}$ to later in this section.

There are several recognisable features of the ${\cal A}(\tau)$ determined from our model: the peak at zero delay is in accord with the physical expectation that a bright image should form at the delay minimum; the parabolic arc seen in figure 3 is responsible for the extended scattering tail stretching out beyond $\tau=100\,{\rm\mu s}$; and the peaks near $\tau=140, 160, 230, 270\,{\rm\mu s}$ in figure 4 correspond to the features labelled ``A, B, C, D'', respectively, by Hill et al (2005) --- these features are seen as discrete intensity concentrations at large delay and positive Doppler-shift in figure 3.

At large negative delays in the top panel of figure 4 we see an intensity level $\sim10^{-5}$, whereas physically we expect zero intensity because wave propagation can only introduce positive group-delays. This is simply due to noise in the reconstruction; note also that a similar floor intensity level is present at large positive delays. The model electric field inevitably includes noise because there is no clear distinction between noise and signal contributions to the input data. By setting the low amplitude coefficients of $\widetilde{U}$ to zero we are able to eliminate much of this intensity floor. Specifically, setting the model coefficients $\tilde{u}_j$ to zero for $|\tilde{u}_j|<0.004$ preserves all of the recognisable signal features in ${\cal A}(\tau)$ but largely removes the noise floor; the result is shown in the lower panel of figure 4.

\begin{figure}
\vskip -1.5cm\hskip -1.5cm\includegraphics[width=11cm]{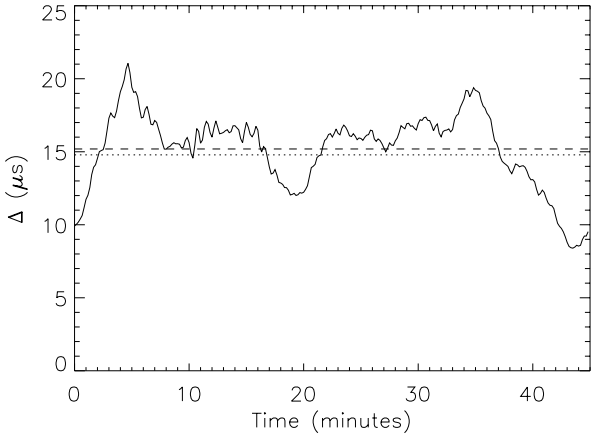}\vskip -6.5cm
\caption{The interstellar propagation delay, $\Delta(t)$ (solid line), as determined from the holographic image shown in figure 3 with low amplitude coefficients ($|\tilde{u}_k|<0.004$) set to zero. Also shown is the (unweighted) mean delay, $\langle\Delta\rangle=15.2\;{\rm \mu s}$ (dashed line), for the observation. The dotted line shows a weighted mean ($14.8\;{\rm \mu s}$), with the weight for each time sample being equal to the intrinsic pulsar flux, $f_k$, as determined by the modelling procedure described in \S2 of this paper.}
\end{figure}

At small negative delays the intensity level in figure 4 is up to an order of magnitude higher than the noise floor. We interpret this as evidence of low-level confusion between the holographic image, $U$, and its complex conjugate, $U^*$ --- as noted in \S3 we expect this confusion to be present at some level. For applications where suppression of the conjugate image is critical it is possible to undertake the holographic image reconstruction entirely in the positive-delay half-space, but we note that this approach is expected to be problematic if the brightest image is not the image with the least delay (WS05).

A useful characterisation of the temporal response of the medium is the intensity-weighted average delay, $\Delta$, defined by
\be
\Delta(t)\equiv\frac{\int_0^\infty\!\!{\rm d}\tau\,\tau\,{\cal I}(\tau,t)}{\int_0^\infty\!\!{\rm d}\tau\,{\cal I}(\tau,t)}.
\ee
The time-average of this quantity, $\langle\Delta\rangle$, provides us with a simple gauge of the influence of wave propagation on pulse arrival time for this line-of-sight for the particular time and frequency intervals covered by our data. For the image $\widetilde{U}$ shown in figure 3 we compute $\langle\Delta\rangle\simeq17\,{\rm\mu s}$. However, this value is clearly an overestimate because noise in the reconstruction -- the noise floor is seen in the top panel of figure 4 -- biases the estimate upward. A better estimate is available if we employ the image $\widetilde{U}$ with low-amplitude coefficients set to zero (as per the lower panel in figure 4); this  yields the result $\langle\Delta\rangle\simeq15.2\,{\rm\mu s}$. Substantial contributions to this mean delay arise from the whole region $0<\tau<120\,{\rm\mu s}$, with relatively minor contributions from each of the discrete features - ``A, B, C, D'' of Hill et al (2005) - seen at large delay; together these features contribute about 10\% of the measured  $\langle\Delta\rangle$. There is no contribution from the artifacts associated with the conjugate image as these occur in the region $\tau<0$ and are excluded by the definition given in equation 6.

Finally we return to the influence of the beat terms, described by ${\cal B}$; these terms cause the  propagation delay $\Delta(t)$ to vary over the course of our 45~minute stretch of data. As noted earlier, the beat terms are not negligible and for our data we find that there are substantial variations in $\Delta$ of $\pm40\%$ around the mean value $\langle\Delta\rangle$; these variations are plotted in figure 5. Large gradients in $\Delta(t)$ are seen in figure 5 so that, for example, a pulse arrival time measurement at the start of our observations and one taken 5 minutes later would have differed by approximately $11\;{\rm \mu s}$. We emphasise that the behaviour seen in figure 5 is specific to this line-of-sight and to the particular set of frequencies covered in our data. We expect that the temporal variations in $\Delta$ would have been smaller if our data had covered a greater bandwidth than the 1.56~MHz used here. With a broader observing bandwidth we would have finer delay resolution in our image $\widetilde{U}$, so we would be better able to separate components which have similar  values of $\tau$ and these separated components would not beat (see equation 4), so the importance of ${\cal B}$ would diminish. It is beyond the scope of this paper to attempt a detailed analysis of delay variations; here we simply note that the type of behaviour seen in figure 5 is potentially problematic for precision pulsar timing.

\section{Discussion}
Over an interval of several months we expect PSR B0834+06 to exhibit changes $\delta\langle\Delta\rangle$ in the mean delay $\langle\Delta\rangle$ as the pulsar moves behind different regions of the scattering screen. The electric field image shown in figure 3 clearly demonstrates that this particular screen has a very patchy distribution of scattering material so that observations made at other epochs are expected to exhibit quite different distributions of power along the parabolic locus. As a hypothetical example we can imagine that at another epoch the holographic image might look like figure 3 at positive Doppler shifts, but show no scattered power at negative Doppler shifts; in this case the mean delay would be roughly half of what we have measured. We therefore expect that  
over an interval of several months PSR~B0834+06 will exhibit changes $\delta\langle\Delta\rangle\sim\langle\Delta\rangle$. The anticipated delay changes $\delta\langle\Delta\rangle\sim15\,{\rm\mu s}$ are very large compared to the precision with which millisecond pulsars can be timed (e.g van~Straten et al 2001). B0834+06 is not a millisecond pulsar, and even if it were it would not be used for precision timing experiments precisely because this line-of-sight is known to exhibit very striking effects from multi-path propagation in the interstellar medium. It is, however, salutary to see how large the influence of the interstellar medium can be on pulse arrival times. Moreover the line-of-sight to B0834+06, although unusual, is not unique --- B1133+16 and B1929+10, for example, appear to show similar, striking effects (Putney and Stinebring 2005). Nor is it a very distant pulsar, so the scattering structures which have been revealed in the present data are probably abundant in the interstellar medium. B0834+06 is, however, a relatively bright source and other, fainter pulsars might be viewed through similar scattering media without being recognised as such because the scattered intensity is small. The data of Putney and Stinebring (2005) support these points as many of the pulsars which they studied appear to show strong scattering arising from {\it multiple,\/}  physically distinct  regions, and they note that very sensitive observations are required in order to reveal the presence of these media. Furthermore the very structured nature of the scattering medium seen in figure 3 tells us that a pulsar which shows no measurable interstellar timing delays at one epoch could exhibit large delays at other epochs. Presumably interstellar scattering media exhibit a broad range of physical properties, with a correspondingly broad range of influence on pulse arrival time measurements, and for any given pulsar we should not ask ourselves ``whether'' but ``at what level'' is an extended scattering tail present?

A commonly used strategy for mitigating the influence of the interstellar medium on pulsar timing experiments is to undertake the timing  observations at high radio frequencies. The rationale for this is based on the assumption that the scattering originates in distributed Kolmogorov turbulence, for which the expected propagation delay falls very rapidly with frequency. Distributed Kolmogorov turbulence is not a good model for what we see in figure 3. Currently we are not able to predict what interstellar timing perturbation would be measured, for this pulsar and this epoch, at radio-frequencies outside the $1.56$~MHz observing band of the present data. It is likely that the interstellar delays would be smaller at higher frequencies -- because the cold plasma refractive index declines with frequency and so the scattering angles decrease at higher frequencies for any given plasma structure -- but we cannot say how much smaller. From data spanning many months, Hemberger and Stinebring (2008) have used a secondary spectrum analysis to estimate multipath propagation delays at frequencies between 1,150~MHz and 1,500~MHz, measuring values which range from one to two orders of magnitude smaller than our estimate of $\langle\Delta\rangle$. Their data refer to a different line-of-sight (PSR B1737+13) and cannot be directly compared with our result; however, their frequency coverage is great enough to allow them to estimate the frequency dependence of the propagation delay for their observations. They find that the propagation delay is consistent with a power-law frequency dependence of the mean propagation delay,  with power-law index $-3.6\pm0.2$ --- slightly less steep than expected for distributed Kolmogorov turbulence (power-law index $-4.4$).

The possibility of large, transient interstellar propagation delays makes it prudent to quantify the interstellar  delays at each epoch where an accurate pulsar timing measurement is desired.  In this paper we have shown how those delays can be quantified when a recording of the pulsar dynamic spectrum is available. The technique we have presented has the merit of being able to accurately determine the relative propagation delays, even for complex scattering geometries, at any epoch of observation. The technique does have its limitations though. Foremost among these is that the delays are all relative measurements and the origin of coordinates (i.e. $\tau=0$) must be determined by some other means; removing this limitation is a high priority goal for future work. 

For cases where the great majority of the scattering arises in a single, thin screen we anticipate that the key to progress lies in constructing the holographic image by modelling the structure of the phase screen, rather than forming the image $\widetilde{U}$ in delay-Doppler space as we have done in this paper. Modelling the phase screen is computationally much more demanding but it has several advantages over the approach we have taken here. First it allows the interstellar propagation delays to be determined on an absolute scale, because the propagation geometry is fixed in the model. Secondly it allows us to calculate the wave field for various different observer locations, thus permitting direct comparisons between different observing epochs and between different telescopes (e.g. Very Long Baseline Interferometry) at the same epoch. Thirdly it permits direct comparison between data at different frequencies. And by the same token there would be no difficulty modelling data taken over a single, wide bandwidth even if the phase screen has large spatial variations in wave dispersion. By contrast the present approach is not ideal for wide bandwidth data because of smearing due to differential dispersive delays, across the band, from a range of dispersion measures.  Finally, if we model the phase screen itself the results immediately provide powerful constraints on models of the scattering medium, because the phase structure tells us the electron column-density structure and the line-of-sight magnetic field if full polarisation information is recorded.

\section{Conclusions}
Interstellar holography is a precise new technique which affords detailed insights into the interstellar propagation of radio pulsar signals. The holographic image reconstructed from our test data reveals a complex scattering structure whose physical nature is unknown at present. Holographic imaging can be used to determine the influence of multi-path propagation on pulse arrival time measurements and thus to correct for these propagation delays. Interstellar propagation delays are unpredictable and can potentially make large contributions  to the systematic errors in pulse arrival time measurements. It is therefore prudent to make provision for holography in every instance where an accurate measurement of the unperturbed arrival time is desired.

\section{Acknowledgments}
Thanks to Don Backer, Barney Rickett and Bill Coles for helpful discussions. 
MAW thanks  R.~Ramachandran for reminding him of the correct terminology for the reconstruction undertaken by WS05. At the University of Sydney this work was supported by the Australian Reseach Council, at the Kapteyn Institute, ASTRON and Leiden by the Netherlands Organisation for Scientific Research, and at Oberlin by the National Science Foundation.

\end{document}